\documentclass{article}
\usepackage[preprint]{spconf}
\copyrightnotice{\copyright IEEE 2022}
\toappear{To appear in \emph{Proc. ICASSP 2022, May 22-27, 2022, Singapore}}
\usepackage{color,xcolor,colortbl}
\usepackage{tabu}
\usepackage{epsfig}
\usepackage{graphicx}
\definecolor{LightCyan}{rgb}{0.88,1,1}

\usepackage{array}
\usepackage{booktabs}
\usepackage{colortbl}
\usepackage{multirow}
\usepackage{float}
\usepackage{caption}
\usepackage[labelformat=simple]{subcaption}

\usepackage{footnote}
\makesavenoteenv{tabular}
\makesavenoteenv{table}

\usepackage{amsmath,amsfonts,amssymb,bm}
\usepackage[super]{nth}

\usepackage{changepage}
\usepackage{extramarks}
\usepackage{fancyhdr}
\usepackage{lastpage}
\usepackage{setspace}
\usepackage{soul}
\usepackage{xspace}

\usepackage{url}
\usepackage{hyperref}
\hypersetup{colorlinks=True, urlcolor=black}
\usepackage[numbers,sort&compress]{natbib}
\usepackage{algorithm, algorithmic}
\usepackage{enumitem}
\usepackage{verbatim}
\usepackage{pifont}
\usepackage[acronyms]{glossaries}
\glsdisablehyper

\newcommand{\sectdot}[1]{Sec.~\ref{sec:#1}}

\newcommand{\ssectdot}[1]{Sec.~\ref{ssec:#1}}

\newcommand{\eqndot}[1]{Eqn.~(\ref{eqn:#1})}

\newcommand{\tbl}[1]{Table~\ref{tab:#1}}

\newcommand{\twosectdot}[2]{Secs.~\ref{sec:#1} and \ref{sec:#2}}

\newcommand{\ignore}[1]{}

\makeatletter
\DeclareRobustCommand\onedot{\futurelet\@let@token\@onedot}
\def\@onedot{\ifx\@let@token.\else.\null\fi\xspace}

\def\ie{\emph{i.e}\onedot}

\def\wrt{w.r.t\onedot}

\makeatother

\definecolor{MyDarkBlue}{rgb}{0,0.08,1}
\definecolor{MyDarkGreen}{rgb}{0.02,0.6,0.02}
\definecolor{MyDarkRed}{rgb}{0.8,0.02,0.02}
\definecolor{MyDarkOrange}{rgb}{0.40,0.2,0.02}
\definecolor{MyPurple}{RGB}{111,0,255}
\definecolor{MyRed}{rgb}{1.0,0.0,0.0}
\definecolor{MyGold}{rgb}{0.75,0.6,0.12}
\definecolor{MyDarkgray}{rgb}{0.66, 0.66, 0.66}

\def\presec{\vspace{-0.2em}}
\def\postsec{\vspace{-0.2em}}
\def\pretbl{\vspace{-0.2em}}
\def\posttbl{\vspace{-0.5em}}

\title{Knowledge Distillation for Neural Transducers\\from Large Self-Supervised Pre-trained Models}
\name{Xiaoyu Yang, Qiujia Li$^*$, Philip C. Woodland\thanks{$^*$Li is funded by a Peterhouse Graduate Studentship.}}
\address{Cambridge University Engineering Dept., Trumpington St., Cambridge, CB2 1PZ U.K.\\
\footnotesize{\texttt{\{xy316,ql264,pcw\}@eng.cam.ac.uk}}\vspace{-1.2em}}
\begin{document}
\ninept
\maketitle

\begin{abstract}
Self-supervised pre-training is an effective approach to leveraging a large amount of unlabelled data to reduce word error rates (WERs) of automatic speech recognition (ASR) systems.
Since it is impractical to use large pre-trained models for many real-world ASR applications,
it is desirable to have a much smaller model while retaining the performance of the pre-trained model.
In this paper, we propose a simple knowledge distillation (KD) loss function for neural transducers that focuses on the one-best path in the output probability lattice under both streaming and non-streaming setups, which allows a small student model to approach the performance of the large pre-trained teacher model.
Experiments on the LibriSpeech dataset show that despite being 10 times smaller than the teacher model, the proposed loss results in relative WER reductions (WERRs) of 11.5\% and 6.8\% on the test-other set for non-streaming and streaming student models compared to the baseline transducers trained without KD using the labelled 100-hour clean data. With an additional 860 hours of unlabelled data for KD, the WERRs increase to 48.2\% and 38.5\% for non-streaming and streaming students. If language model shallow fusion is used for producing distillation targets, a further improvement in the student model is observed.

\end{abstract}

\presec
\section{Introduction}
\postsec
Self-supervised learning has emerged as a paradigm to learn general data representations. The model is first pre-trained on a very large amount of unlabelled data and then fine-tuned with task-specific labelled data.
The effectiveness of this approach was first demonstrated in various natural language processing tasks~\cite{devlin2018bert,brown2020GPT3} and, more recently, in automatic speech recognition (ASR)~\cite{baevski2019vqw2v2,baevski2020wav2vec}.
Promisingly, some of these models could reach state-or-the-art performance without relying on a large amount of labelled data.
Although self-supervised pre-training can give excellent performance for ASR, the models are typically very large with hundreds of millions or even billions of parameters. This leads to large memory requirements and a long inference time, which hinders the direct deployment of such models for real-world ASR applications~\cite{peng2021shrinking}. Moreover, the pre-trained speech encoder is generally bidirectional or attends to a very wide range of past and future context~\cite{baevski2020wav2vec,baevski2019vqw2v2}, which is problematic for ASR applications that require streaming processing of speech data.

To address this issue, various model compression methods have been proposed to reduce the model size dramatically with only a moderate increase in the WER. 
For example, singular value decomposition of weight matrices can reduce the number of parameters in acoustic models using feed-forward networks~\cite{xue2013restructuring} and recurrent neural networks (RNNs)~\cite{prabhavalkar2016compression} and a sparse pruning method can shrink the model size by setting weights with small values to zero~\cite{zhen2021sparsification}.
Instead of direct manipulation of model parameters, knowledge distillation (KD)~\cite{hinton2015knowledgedistillation} is an alternative approach for model compression, where a small student model learns to mimic the output behaviour of a large teacher model. As KD only depends on the outputs from the teacher and student models, the student model is free to have a different representation of the input data and also a different model architecture. KD is also able to take advantage of unlabelled data as only the output from the teacher model is needed to train the student model~\cite{cao2021improving, doutre2021improving}. For conventional hidden Markov model-based systems, KD has been applied to the acoustic model at the frame level~\cite{Li2014LearningSD} and the sequence level~\cite{Wong2016SequenceST}. For end-to-end (E2E) ASR, KD has been applied for connectionist temporal classification (CTC) models~\cite{takashima2018ctcseqkd, kurata2018improvedCTCkd, takashima2019investigationCTCseqkd,Yoon2021TutorNetTF} and attention-based encoder-decoder models~\cite{2018CompressionLAS,Yoon2021TutorNetTF} by either matching the hidden representations or the softmax output distributions between the teacher and student models. The neural transducer~\cite{graves2012RNNT} is another widely adopted E2E ASR model because of its competitive performance and readiness for streaming applications~\cite{rao2017exploring, he2019streaming}. However, applying KD to neural transducers is less straightforward as computing the distillation loss on the whole output probability lattice from transducers is very expensive and memory-inefficient, especially for long sequences. \cite{doutre2021improving} trained the student transducer model directly from the ``pseudo transcriptions'', \ie erroneous recognition results from the teacher model, instead of using the output distributions. \cite{panchapagesan2021fulllatticernntkd} carried out KD for transducers using a collapsed version of the full output distributions to reduce the memory requirement during training. As a workaround, \cite{swaminathan2021codert} performs KD over the encoder hidden representations.

In this paper, we investigate how to effectively distil knowledge from a large pre-trained transducer model to a much smaller one with a limited amount of labelled data. Instead of using the collapsed distribution over the whole probability lattice~\cite{panchapagesan2021fulllatticernntkd}, using the full distribution of the one-best path in the output probability lattice yields better performance. The proposed distillation loss can be simply adapted to transfer knowledge from a non-streaming teacher to a streaming student by introducing a time-shift variable that controls the amount of future context seen by the encoder.
In our experiments, the teacher is a wav2vec 2.0 model~\cite{baevski2020wav2vec} with the waveform as input, whereas the student is a Conformer model~\cite{gulati2020conformer} with filter bank features as input. Although the student model is 10 times smaller than the teacher model, experiments on LibriSpeech show that the proposed method significantly improves the student's performance for both non-streaming and streaming cases. Further word error rate (WER) reduction is obtained by using unlabelled data for KD or augmenting the distillation targets by a language model (LM).

In the rest of this paper, \sectdot{background} briefly reviews self-supervised pre-training for ASR and the neural transducer models. In \sectdot{method}, the proposed distillation loss for streaming and non-streaming transducers are introduced.
The experimental setup and results are described in \twosectdot{setup}{results}. Conclusions are drawn in \sectdot{conclusion}.
\presec
\section{Self-supervised Pre-training for E2E ASR}
\postsec
\label{sec:background}


\subsection{wav2vec 2.0 for Self-Supervised Pre-training}
\postsec
\label{sec:w2v}
Self-supervised learning for E2E ASR has shown promising results in the last few years~\cite{baevski2019vqw2v2, jiang2019improving, baevski2020wav2vec}. Among these methods, wav2vec 2.0 (w2v2) is one of the most effective frameworks.
The raw speech waveform $\mathbf{w}$ is first processed by a convolutional neural network feature extractor $\mathcal{M}$ to generate latent speech representations $\mathcal{M}(\mathbf{w})=\mathbf{z}_{1:T}$, which are then sent to a Transformer network $\mathcal{N}$ to generate contextualised speech representations $\mathcal{N}(\mathbf{z}_{1:T}) = \mathbf{c}_{1:T}$. A quantisation module $\mathcal{Q}$ discretises the latent speech representation to $\mathbf{q}_t$.
Pre-training is carried out by predicting the correct quantised representations of the masked latent speech representations from a set of distractors.
This training objective for each $\mathbf{z}_t$ can be expressed with a contrastive loss function \eqndot{w2v2}, where $Q_t$ is a set consisting of the correct quantisation $\mathbf{q}_t$ and other distractors, $\kappa$ is the temperature and $\textsc{sim}$ denotes a similarity measure. An additional quantisation diversity loss is added to construct the final loss during pre-training.
\begin{align}
    \mathcal{L}_{\text{contrastive}} = -\log \frac{\exp(\textsc{sim}(\mathbf{c}_t, \mathbf{q}_t)/\kappa)}{\sum_{\mathbf{q}\in Q_t}\exp(\textsc{sim}(\mathbf{c}_t,\mathbf{q})/\kappa)}.
    \label{eqn:w2v2}
\end{align}

The pre-trained model is used to initialise the encoder of an E2E ASR model. The encoder is then fine-tuned together with additional output layers or decoders using labelled data for ASR. Although the original pre-trained models are fine-tuned with the CTC objective~\cite{baevski2020wav2vec}, they can also be used to initialise transducer models~\cite{Zhang2020PushingTL}.

\presec
\subsection{Neural Transducer Model}
\postsec
\label{sec:rnnt}
CTC~\cite{graves2006CTC} is one of the earliest frameworks for E2E ASR. Since the CTC assumption that output tokens at different timestamps are conditionally independent is not realistic for ASR, the RNN transducer was proposed to address this limitation~\cite{graves2012RNNT}. More recently, RNNs in the transducer encoder have  been replaced in order to give improved performance and leading to Transformer transducers~\cite{Zhang2020TransformerTA} and Conformer transducers~\cite{gulati2020conformer}.
A neural transducer model consists of an encoder $\mathcal{F}$, a prediction network $\mathcal{G}$ and a joint network $\mathcal{J}$. During training, given an input feature sequence $\mathbf{X} = \mathbf{x}_{1:T}$ of length $T$, the encoder $\mathcal{F}$ extracts the latent representations $\mathbf{f}_{1:T}$. The prediction network predicts the next symbol $y_{u}$ given all previous symbols ${y_{1:u-1}}$ for $u=1,\dots,U$, similar to a language model. The output from the prediction network $\mathcal{G}(\mathbf{y}) = \mathbf{g}_{1:U}$ is then fed to $\mathcal{J}$ together with $\mathbf{f}_{1:T}$ to generate an output distribution lattice $\mathbf{Z}$ of dimension $K \times T\times U$, where $K$ is the output vocabulary size. The probability $p(k|t,u)$ of emitting token $k$ at node $(t,u)$ in $\mathbf{Z}$ is then defined as the $k$-th entry in the vector obtained after applying softmax to $\mathbf{Z}(t,u)$. Neural transducer training aims to maximise the probability $p(\mathbf{y}|\mathbf{X})$, which can be efficiently computed using a forward-backward algorithm on the lattice $\mathbf{Z}$~\cite{graves2012RNNT}.

Neural transducers generally outperform CTC models due to their ability to model the dependency across output tokens. As the decoding procedure is frame-synchronous, transducers are suitable for streaming applications if the encoder has only a limited exposure to future acoustic context~\cite{rao2017exploring, he2019streaming}.
\presec
\section{Knowledge Distillation for Transducers}
\postsec
\label{sec:method}

Knowledge distillation (KD)~\cite{hinton2015knowledgedistillation} is a widely used model compression technique that trains a small student model to match the output of a large teacher model.
For a classification task with $K$ classes, KD usually minimises the Kullback–Leibler (KL) divergence between the teacher output distribution $p_{\mathcal{T}}$ and the student output distribution $p_{\mathcal{S}}$ as given in \eqndot{KL-divergence}.
\begin{align}
    \mathcal{L}_{\text{KD}}=-\sum_{k=1}^{K} p_{\mathcal{T}}(k) \log \frac{p_{\mathcal{S}}(k)}{p_{\mathcal{T}}(k)}.
    \label{eqn:KL-divergence}
\end{align}

Since it does not require ground truth labels, KD can leverage unlabelled data for student model training. KD also provides greater flexibility for model compression since it does not constrain the input features or the model architectures of the teacher and student models to be the same.

\presec
\presec
\subsection{KD for Non-Streaming Transducer}
\postsec
\label{ssec:non-streaming}
As described in \sectdot{rnnt}, a neural transducer generates an output distribution lattice $\mathbf{Z}$ from an acoustic feature sequence $\mathbf{X}$ and a label sequence $\mathbf{y}$, from which the total conditional probability of the label sequence given the acoustic sequence can be computed by summing over all possible alignments. Therefore, the output lattice of a teacher transducer is an obvious distillation target for the student model. A straightforward distillation objective is to minimise the KL divergence (or cross-entropy as the teacher labels are fixed) between the output distribution lattices of the student $\mathbf{Z}_{\mathcal{S}}$ and teacher $\mathbf{Z}_{\mathcal{T}}$, which results in the loss function in \eqndot{fulllatticeKD}. As a result, the output distribution of the student model should imitate the teacher distribution over the entire lattice. However, this method is impractical due to its large memory and computation complexities of $\mathcal{O}(KTU)$.
\begin{align}
    \mathcal{L}_{\text{KD}} = -\sum_{t=1}^{T}\sum_{u=1}^{U}\sum_{k=1}^{K} \mathbf{Z}_{\mathcal{T}}(k,t,u) \log \mathbf{Z}_{\mathcal{S}}(k,t,u).
    \label{eqn:fulllatticeKD}
\end{align}

To reduce the computational cost of KD for transducers, distillation based on a collapsed distribution of all nodes in the lattice was proposed~\cite{panchapagesan2021fulllatticernntkd}. By only considering the blank symbol, the correct output symbol and all the remaining symbols, the output distribution for each node is reduced from $K$ to only 3 classes. The complexity becomes $\mathcal{O}(TU)$.
However, this method ignores the correlation across different output symbols compared to \eqndot{fulllatticeKD}, which could be important for training the student model using KD.

To achieve memory efficiency and also preserve the full distribution, we propose to distil knowledge only over the one-best path in the lattice instead of over the whole lattice. It has been shown that only a small proportion of nodes in the lattice contribute to the final decoding results~\cite{jain2019rnntconstrainedbm}, which indicates that a large number of paths in the lattice have low probabilities.
As an approximation, only the nodes along the one-best path are considered as important. The distillation loss for one-best path in the output distribution lattice is
\begin{align}
    \mathcal{L}_{\text{KD}} \approx -\sum_{(t,u)\in \textsc{1Best}}\sum_{k=1}^{K} \mathbf{Z}_{\mathcal{T}}(k,t,u) \log \mathbf{Z}_{\mathcal{S}}(k,t,u),
    \label{eqn:rnnt onebest}
\end{align}
where $\textsc{1Best}$ is the most likely alignment between $\mathbf{X}$ and $\mathbf{y}$ containing the blank symbol. For an input utterance of length $T$ and a target sequence of length $U$, the memory complexity is only $\mathcal{O}(K(T+U))$. Compared with \cite{panchapagesan2021fulllatticernntkd}, the proposed approach preserves the full distribution for each node at the expense of ignoring other possible alignments. Assuming the output lattice is relatively sparse, the proposed method is expected to transfer richer information than the collapsed distribution~\cite{panchapagesan2021fulllatticernntkd}. Furthermore, using the one-best alignment allows the alignment to be delayed when training a streaming student from a non-streaming teacher as discussed in \ssectdot{streaming}.

\presec
\subsection{KD for Streaming Transducer}
\postsec
\label{ssec:streaming}
In contrast to non-streaming transducers, directly applying one-best KD as in \eqndot{rnnt onebest} to streaming transducers may not be sensible. As little (or no) future context is available for a streaming transducer, it tends to emit symbols later than non-streaming models to gather more future information~\cite{kurata2020knowledge}.
In other words, the most likely alignment for a streaming transducer normally lags behind its non-streaming counterpart by several time steps.
As the pre-trained teacher model is non-streaming, directly applying \eqndot{rnnt onebest} to perform KD on a streaming student model may force the student model to ``guess'' into the future, which may lead to poor performance.
To this end, a hyperparameter $\tau$ is introduced to delay the non-streaming alignment, which allows the student model to emit symbols $\tau$ frames later than its teacher. $\tau$ can be tuned to find the best trade-off between performance and latency. To train a streaming student model, the modified distillation loss is
\begin{align}
    \mathcal{L}_{\text{KD}} \approx -\sum_{(t,u)\in \textsc{1Best}}\sum_{k=1}^{K} \mathbf{Z}_{\mathcal{T}}(k,t,u) \log \mathbf{Z}_{\mathcal{S}}(k,t+\tau,u).
    \label{eqn:rnnt onebest streaming}
\end{align}

\presec
\subsection{Using Unlabelled Data and LM Fusion for KD}
\postsec
Unlike other model compression techniques, KD is a data-driven approach that is able to leverage unlabelled data to further improve the student model. For unlabelled speech data, the teacher model is used to generate one-best hypotheses via beam search. The one-best alignment and the output probabilities of each node on it for each utterance are used as distillation targets.

During decoding, an external LM can be used to improve the quality of the transcription~\cite{bahdanau2016end, kannan2018analysis}, which is beneficial to KD training. 
In this work, log-linear interpolation (shallow fusion)~\cite{bahdanau2016end} of the transducer and an external LM is adopted to construct distillation targets containing LM information,
\begin{align}
    \mathbf{Z}'_{\mathcal{T}}(t, u) = \textsc{Softmax}\big(\log(\mathbf{Z}_{\mathcal{T}}(t,u))+\beta\log(\textsc{LM}(t))\big),
\end{align}
where $\textsc{LM}(t)$ is a $K$-dimensional vector representing the LM output distribution given the previously predicted sequence before $t$ and $\beta$ is a tuned LM weight. Note that the LM score for the blank symbol is set to zero when the blank symbol is emitted and $\min(\text{LM}(t))$ when a non-blank symbol is emitted. 
For better convergence, the distillation loss is interpolated with the original transducer loss,
\begin{align}
    \mathcal{L} = \mathcal{L}_{\text{transducer}} + \lambda \mathcal{L}_{\text{KD}},
    \label{eqn:kd full loss}
\end{align}
where $\lambda$ is the interpolation coefficient.

\presec
\section{Experimental Setup}
\postsec
\label{sec:setup}

\subsection{Model}
\postsec
For the teacher model, the base w2v2 model~\cite{baevski2020wav2vec} was used to initialise the encoder of a transducer model.
The original w2v2 model takes the waveform as input and has an output frequency of 100 Hz, whereas filter bank features are more commonly used as ASR input and operate at 50 Hz. To achieve efficient knowledge distillation, a sub-sampling layer was added on top of the pre-trained model to reduce the frequency of the encoder output to 50 Hz. 
The student model is a small Conformer transducer model~\cite{gulati2020conformer}. Both models have a single-layer long short-term memory (LSTM) projection network. The output vocabulary has 256 subword units generated using SentencePiece~\cite{kudo2018sentencepiece}. The details of both models are given in \tbl{model comparison}. Note that the student model has more than 10 times fewer parameters than the teacher model.
For the LM used for shallow fusion, a 2-layer LSTM with 2048 units was trained from the LibriSpeech LM corpus. The vocabularies of the transducer models and the LM are the same. All models were implemented in ESPnet~\cite{watanabe2018espnet}.
\begin{table}[ht]
    \pretbl
    \centering
    \begin{tabular}{ccc}
        \toprule
                & Teacher model & Student model \\
        \midrule
        Encoder & w2v2~\cite{baevski2020wav2vec} & Conformer-S~\cite{gulati2020conformer} \\
        Encoder dimension & 768 & 144\\
        Decoder dimension & 640 & 320 \\
        Number of parameters & 99.2M & 9.7M\\
        \bottomrule
    \end{tabular}
    \caption{Details of the teacher and student transducer models.}
    \label{tab:model comparison}
    \posttbl
    \posttbl
\end{table}

\presec
\subsection{Data}
\postsec
The LibriSpeech dataset~\cite{Panayotov2015LibrispeechAA} was used for the experiments. The full training set has 960 hours of audiobook recordings. Among these, ``train-clean-100'' was used as labelled data while the remaining 860 hours were treated as unlabelled data. The teacher model was pre-trained using the full training set and fine-tuned on the 100-hour labelled subset following \cite{baevski2020wav2vec}.
\presec
\section{Experimental Results}
\postsec
\label{sec:results}

\subsection{Baselines}
\postsec
\tbl{ref models} gives the word error rate (WER) of both the fine-tuned w2v2 models and the baseline conformer models trained on the LibriSpeech 100-hour subset. The fine-tuning configuration of the teacher model follows \cite{baevski2020wav2vec}. Although our implementation shares the same pre-trained encoder as the \textsc{Base} model in \cite{baevski2020wav2vec},
the performance is better because our fine-tuned w2v2 model is a transducer model with BPE modelling units instead of a CTC model with grapheme units~\cite{baevski2020wav2vec}. For the baseline Conformer-S models, SpecAugment~\cite{park2019specaugment} was used during training employing 2 frequency masks with $F=27$ and 10 time masks with a maximum time-mask ratio of 0.05.
\begin{table}[ht]
    \pretbl
    \centering
    \begin{tabular}{lrrrr}
        \toprule
        \multirow{2}{*}{Models}    & \multicolumn{2}{c}{dev}   & \multicolumn{2}{c}{test}  \\
        \cmidrule(lr){2-3}  \cmidrule(lr){4-5}
        & \multicolumn{1}{c}{clean} & \multicolumn{1}{c}{other} & \multicolumn{1}{c}{clean} & \multicolumn{1}{c}{other} \\ 
        \midrule
        w2v2 CTC (\cite{baevski2020wav2vec})   &  6.1 & 13.5 &  6.1 & 13.3\\
        w2v2 transducer (ours)                        &  5.1 & 12.2  &  5.2 & 11.8 \\
        \;\;\;\; + LM shallow fusion                   &  4.2 & 10.4 &  4.3 & 10.1\\
        \midrule
        Conformer-S (non-streaming)              & 7.5 & 20.9 & 8.0 & 21.8\\
        Conformer-S (streaming)                  & 11.2 & 28.4 & 12.2 & 29.6\\
        \bottomrule
    \end{tabular}
    \caption{WERs of the w2v2 models and the baseline small Conformer models trained on LibriSpeech 100-hour subset. Our fine-tuned w2v2 transducer will be used as the teacher model.}
    \label{tab:ref models}
    \posttbl
\end{table}

As expected, the w2v2 transducer gives much lower WERs than the non-streaming Conformer-S model as it benefits from both self-supervised pre-training and a larger model size. For the streaming transducer, an attention mask was added to remove the contribution of future frames in the encoder and the 1D-convolution changed to 1D causal convolution to only attend to previous frames. Consequently, the WER for the streaming Conformer-S increases.

In the following experiments, the w2v2 transducer (2nd row in \tbl{ref models}) was used as the teacher model. The student transducer model with the same architecture as the Conformer-S tries to approach the performance of the teacher model. For both the non-streaming and the streaming setups, the 100-hour labelled training data was used first to verify the proposed KD approach. Then, the remaining 860-hour training data was used as unlabelled data to further improve the student models.

\presec
\subsection{KD for Non-Streaming Transducers}
\postsec
\label{ssec:non-streaming_trans}
Two KD training strategies were investigated for the proposed distillation loss. The first strategy (ST1) trains the student transducer model from scratch using \eqndot{kd full loss}. The second strategy (ST2) initialises the student model from a model trained on ground truth or pseudo transcriptions and then uses the proposed loss in \eqndot{kd full loss} for fine-tuning.
\begin{table}[ht]
    \pretbl
    \centering
    \begin{tabular}{lrrrr}
        \toprule
        \multirow{2}{*}{Transducer models}    & \multicolumn{2}{c}{dev}   & \multicolumn{2}{c}{test}  \\
        \cmidrule(lr){2-3}  \cmidrule(lr){4-5}
        & \multicolumn{1}{c}{clean} & \multicolumn{1}{c}{other} & \multicolumn{1}{c}{clean} & \multicolumn{1}{c}{other} \\ 
        \midrule
        \multicolumn{5}{l}{\textbf{Reference models: 100h labelled}}\\
        w2v2 transducer (teacher) & 5.1 & 12.2 & 5.2 & 11.8\\
        Conformer-S (baseline)    & 7.5 & 20.9 & 8.0 & 21.8\\
        \midrule
        \multicolumn{5}{l}{\textbf{Student models: 100h labelled}}\\
        $\lambda=0.001$, collapsed \cite{panchapagesan2021fulllatticernntkd} & 7.1 & 20.7 & 7.5 & 20.9\\
        $\lambda=0.001$, collapsed \cite{panchapagesan2021fulllatticernntkd}, ST2 & 6.8 & 19.2 & 7.2 & 19.8\\
        $\lambda=0.1$, ST1       & 7.1 & 19.8 & 7.6 & 20.5 \\
        $\lambda=0.1$, ST2       & 6.7 & 19.0 & 7.1 & 19.3 \\
        $\lambda=0.1$, ST2 [+LM] & 6.7 & 18.9 & 7.0 & 19.3 \\
        \midrule
        \multicolumn{5}{l}{\textbf{Student models: 100h labelled + 860h unlabelled}}\\
        $\lambda=0.0$            & 5.5 & 11.5 & 5.6 & 11.9 \\
        $\lambda=0.0$ [+LM]      & 4.7 & 10.4 & 4.8 & 10.9 \\
        $\lambda=0.1$, ST1       & 5.5 & 11.3 & 5.5 & 11.7 \\
        $\lambda=0.1$, ST2       & 5.3 & 10.9 & 5.4 & 11.3 \\
        $\lambda=0.1$, ST2 [+LM] & 4.6 & 10.1 & 4.8 & 10.4 \\
        \bottomrule
    \end{tabular}
    \caption{WERs of non-streaming transducer models. $\lambda$ is the weight of KD loss. [+LM] means either the teacher output distributions or the pseudo transcriptions are obtained with LM shallow fusion.}
    \label{tab:non-streaming KD}
    \posttbl
\end{table}

Five key observations can be made from \tbl{non-streaming KD}.
First, for student models trained from 100h of labelled data, the proposed distillation loss using the full distribution of the one-best alignment slightly outperforms KD using a collapsed distribution on the whole lattice~\cite{panchapagesan2021fulllatticernntkd}.
Second, for student models trained from 100h labelled data and 860h unlabelled data, the proposed KD method yields lower WERs than simply using the pseudo transcriptions ($\lambda=0.0$) of the unlabelled data because the output distribution carries more information from the teacher to the student than the pseudo transcriptions.
Third, ST2 has lower WERs than ST1 for both the 100h and the 960h setups. As a result, ST2 will be used for the streaming experiments in \ssectdot{streaming_trans}.
Fourth, the relative WER reductions (WERRs) on test-clean and test-other are 11.3\% and 11.5\% from using the proposed KD loss compared with the baseline. The WERRs increase to 32.5\% and 48.2\% when the additional 860h of unlabelled data was used for KD.
Lastly, to distil knowledge from both the teacher model and the LM together, shallow fusion was used to generate the pseudo transcriptions on the unlabelled data and augment the output distributions. This leads to a WERR of 12.5\% for the 100h setup compared with the baseline on test-clean. After incorporating the 860h of unlabelled data, the WERRs on test-clean and test-other improve to 40.0\% and 52.3\%.

\presec
\subsection{KD for Streaming Transducers}
\postsec
\label{ssec:streaming_trans}
By introducing the parameter $\tau$ for delaying the alignment, the proposed one-best distillation loss was applied to train streaming student models.
As before, the distillation targets were still generated from the non-streaming teacher w2v2 transducer. When the streaming student model was trained to match the alignment of the non-streaming teacher, \ie when $\tau=0$, the WER of the student model is even higher than the baseline due to a lack of future context. After observing the one-best alignment of a streaming transducer can be a number of steps behind the one-best alignment of a non-streaming transducer, the WERs for $\tau=6$ and $\tau=7$ (equivalent to a look-ahead of 240 ms and 280 ms) are shown in \tbl{streaming KD}. Although more delay steps result in improved WERs, the latency of the ASR model also increases. Therefore, setting a sensible value of $\tau$ is key to training a student model using KD.
Compared to the streaming baseline, the proposed distillation approach achieves WERRs of 7.4\% and 6.8\% on test-clean and test-other in 100h experiments. With the additional 860h unlabelled data, the student model trained with KD outperforms the one trained only with pseudo transcriptions and increases WERRs to 19.7\% and 38.5\% \wrt the streaming baseline.
\begin{table}[ht]
    \pretbl
    \centering
    \begin{tabular}{lrrrr}
        \toprule
        \multirow{2}{*}{Transducer models}    & \multicolumn{2}{c}{dev}   & \multicolumn{2}{c}{test}  \\
        \cmidrule(lr){2-3}  \cmidrule(lr){4-5}
        & \multicolumn{1}{c}{clean} & \multicolumn{1}{c}{other} & \multicolumn{1}{c}{clean} & \multicolumn{1}{c}{other} \\ 
        \midrule
        \multicolumn{5}{l}{\textbf{Reference models: 100h labelled}}\\
        w2v2 transducer (teacher) &  5.1 & 12.2 &  5.2 & 11.8\\
        Conformer-S (baseline)         & 11.2 & 28.4 & 12.2 & 29.6\\
        \midrule
        \multicolumn{5}{l}{\textbf{Student models: 100h labelled}}\\
        $\lambda=0.1$, $\tau$=0, ST2 & 20.2 & 39.7& 20.5 & 41.4 \\
        $\lambda=0.1$, $\tau$=6, ST2 & 10.7 & 26.7& 11.3 & 28.0 \\
        $\lambda=0.1$, $\tau$=7, ST2 & 10.6 & 26.4& 11.3 & 27.6\\
        \midrule
        \multicolumn{5}{l}{\textbf{Student models: 100h labelled + 860h unlabelled}}\\
        $\lambda=0.0$                & 9.5  & 19.3 & 10.6 & 19.7  \\
        $\lambda=0.1$, $\tau$=7, ST2 & 9.2  & 18.3 & 9.8  & 18.2 \\
        \bottomrule
    \end{tabular}
    \caption{WERs of streaming transducer models. $\lambda$ is the weight of the KD loss. $\tau$ is the number of delayed steps for the one-best alignment from the teacher model.}
    \label{tab:streaming KD}
    \posttbl
\end{table}


\presec
\vspace{-0.2cm}
\section{Conclusions}
\postsec
\label{sec:conclusion}
In this paper, we propose a simple and efficient knowledge distillation (KD) loss for neural transducers using the full distribution along the one-best alignment of the output distribution lattice. This approach can be easily extended to distil knowledge from a non-streaming transducer to a streaming one. We showed the effectiveness of the proposed approach by distilling knowledge from a non-streaming teacher model, initialised from a large self-supervised pre-trained model, to both non-steaming and streaming student models that are more than 10 times smaller. Experiments also show that the performance of the student model can be greatly improved by having more unlabelled data for KD or augmenting the distillation targets by language model shallow fusion.
With better teacher models or more unlabelled data, the student models would be expected to reach better performance. In future, we will explore how to best transfer information from an external language model to the student model, and how to distil knowledge more effectively from a non-streaming teacher to a streaming student.



\newpage
\section{References}
\vspace{-0.5em}
\begingroup
\renewcommand{\section}[2]{}
\bibliographystyle{IEEEbib}
\bibliography{refs}

\begin{thebibliography}{10}

\bibitem{devlin2018bert}
J.~Devlin, M.W.~Chang, K.~Lee, \& K.~Toutanova,
\newblock ``BERT: Pre-training of deep bidirectional transformers for language
  understanding,''
\newblock {\em Proc. }{\em NAACL}, Minneapolis, 2019.

\bibitem{brown2020GPT3}
T.~Brown, B.~Mann, N.~Ryder, M.~Subbiah, J.D.~Kaplan, \& ~Dhariwal,
\newblock ``Language models are few-shot learners,''
\newblock {\em Proc. }{\em NeurIPS}, Vancouver, 2020.

\bibitem{baevski2019vqw2v2}
A.~Baevski, S.~Schneider, \& M.~Auli,
\newblock ``vq-wav2vec: Self-supervised learning of discrete speech
  representations,''
\newblock {\em Proc. }{\em ICLR}, New Orleans, 2019.

\bibitem{baevski2020wav2vec}
A.~Baevski, Y.~Zhou, A.~Mohamed, \& M.~Auli,
\newblock ``wav2vec 2.0: A framework for self-supervised learning of speech
  representations,''
\newblock {\em Proc. }{\em NeurIPS}, Vancouver, 2020.

\bibitem{peng2021shrinking}
Z.~Peng, A.~Budhkar, I.~Tuil, J.~Levy, P.~Sobhani, R.~Cohen, \& J.~Nassour,
\newblock ``Shrinking bigfoot: Reducing wav2vec 2.0 footprint,''
\newblock {\em arXiv preprint arXiv:2103.15760}, 2021.

\bibitem{xue2013restructuring}
J.~Xue, J.~Li, \& Y.~Gong,
\newblock ``Restructuring of deep neural network acoustic models with singular
  value decomposition,''
\newblock {\em Proc. }{\em Interspeech}, Lyon, 2013.

\bibitem{prabhavalkar2016compression}
R.~Prabhavalkar, O.~Alsharif, A.~Bruguier, \& I.~McGraw,
\newblock ``On the compression of recurrent neural networks with an application
  to LVCSR acoustic modeling for embedded speech recognition,''
\newblock {\em Proc. }{\em ICASSP}, Shanghai, 2016.

\bibitem{zhen2021sparsification}
K.~Zhen, H.D.~Nguyen, F.J.~Chang, A.~Mouchtaris, \& A.~Rastrow,
\newblock ``Sparsification via compressed sensing for automatic speech
  recognition,''
\newblock {\em Proc. }{\em ICASSP}, Toronto, 2021.

\bibitem{hinton2015knowledgedistillation}
G.~Hinton, O.~Vinyals, \& J.~Dean,
\newblock ``Distilling the knowledge in a neural network,''
\newblock {\em Proc. }{\em NIPS Deep Learning Workshop}, Montreal, 2014.

\bibitem{cao2021improving}
S.~Cao, Y.~Kang, Y.~Fu, X.~Xu, S.~Sun, \etal,
\newblock ``Improving streaming transformer based ASR under a framework of
  self-supervised learning,''
\newblock {\em Proc. }{\em Interspeech}, Brno, 2021.

\bibitem{doutre2021improving}
T.~Doutre, W.~Han, M.~Ma, Z.~Lu, C.C.~Chiu, \etal,
\newblock ``Improving streaming automatic speech recognition with non-streaming
  model distillation on unsupervised data,''
\newblock {\em Proc. }{\em ICASSP}, Toronto, 2021.

\bibitem{Li2014LearningSD}
J.~Li, R.~Zhao, J.T.~Huang, \& Y.~Gong,
\newblock ``Learning small-size DNN with output-distribution-based criteria,''
\newblock {\em Proc. }{\em Interspeech}, Singapore, 2014.

\bibitem{Wong2016SequenceST}
J.H.M.~Wong \& M.J.F.~Gales,
\newblock ``Sequence student-teacher training of deep neural networks,''
\newblock {\em Proc. }{\em Interspeech}, San Francisco, 2016.

\bibitem{takashima2018ctcseqkd}
R.~Takashima, S.~Li, \& H.~Kawai,
\newblock ``An investigation of a knowledge distillation method for CTC
  acoustic models,''
\newblock {\em Proc. }{\em ICASSP}, Alberta, 2018.

\bibitem{kurata2018improvedCTCkd}
G.~Kurata \& K.~Audhkhasi,
\newblock ``Improved knowledge distillation from bi-directional to
  uni-directional LSTM CTC for end-to-end speech recognition,''
\newblock {\em Proc. }{\em SLT}, Athens, 2018.

\bibitem{takashima2019investigationCTCseqkd}
R.~Takashima, L.~Sheng, \& H.~Kawai,
\newblock ``Investigation of sequence-level knowledge distillation methods for
  CTC acoustic models,''
\newblock {\em Proc. }{\em ICASSP}, Brighton, 2019.

\bibitem{Yoon2021TutorNetTF}
J.~Yoon, H.~Lee, H.Y.~Kim, W.I.~Cho, \& N.S.~Kim,
\newblock ``TutorNet: Towards flexible knowledge distillation for end-to-end
  speech recognition,''
\newblock {\em IEEE/ACM Trans. on Audio, Speech, and Language Processing}, vol.
  29, 2021.

\bibitem{2018CompressionLAS}
R.~Pang, T.~Sainath, R.~Prabhavalkar, S.~Gupta, \& C.C.~Chiu,
\newblock ``Compression of end-to-end models,''
\newblock {\em Proc. }{\em Interspeech}, Hyderabad, 2018.

\bibitem{graves2012RNNT}
A.~Graves,
\newblock ``Sequence transduction with recurrent neural networks,''
\newblock {\em Proc. }{\em ICML Workshop on Representation Learning},
  Edinburgh, 2012.

\bibitem{rao2017exploring}
K.~Rao, H.~Sak, \& R.~Prabhavalkar,
\newblock ``Exploring architectures, data and units for streaming end-to-end
  speech recognition with RNN-transducer,''
\newblock {\em Proc. }{\em ASRU}, Okinawa, 2017.

\bibitem{he2019streaming}
Y.~He, T.N.~Sainath, R.~Prabhavalkar, I.~McGraw, R.~Alvarez, \etal,
\newblock ``Streaming end-to-end speech recognition for mobile devices,''
\newblock {\em Proc. }{\em ICASSP}, Brighton, 2019.

\bibitem{panchapagesan2021fulllatticernntkd}
S.~Panchapagesan, D.S.~Park, C.C.~Chiu, Y.~Shangguan, Q.~Liang, \&
  A.~Gruenstein,
\newblock ``Efficient knowledge distillation for RNN-transducer models,''
\newblock {\em Proc. }{\em ICASSP}, Toronto, 2021.

\bibitem{swaminathan2021codert}
R.V.~Swaminathan, B.~King, G.P.~Strimel, J.~Droppo, \& A.~Mouchtaris,
\newblock ``CoDERT: Distilling encoder representations with co-learning for
  transducer-based speech recognition,''
\newblock {\em Proc. }{\em Interspeech}, Brno, 2021.

\bibitem{gulati2020conformer}
A.~Gulati, J.~Qin, C.C.~Chiu, N.~Parmar, Y.~Zhang, \etal,
\newblock ``Conformer: Convolution-augmented transformer for speech
  recognition,''
\newblock {\em Proc. }{\em Interspeech}, Shanghai, 2020.

\bibitem{jiang2019improving}
D.~Jiang, X.~Lei, W.~Li, N.~Luo, Y.~Hu, \etal,
\newblock ``Improving transformer-based speech recognition using unsupervised
  pre-training,''
\newblock {\em Proc. }{\em Interspeech}, Graz, 2019.

\bibitem{Zhang2020PushingTL}
Y.~Zhang, J.~Qin, D.S.~Park, W.~Han, C.~Chiu, \etal,
\newblock ``Pushing the limits of semi-supervised learning for automatic speech
  recognition,''
\newblock {\em Proc. }{\em NeurIPS SAS Workshop}, Vancouver, 2020.

\bibitem{graves2006CTC}
A.~Graves, S.~Fern{\'a}ndez, F.~Gomez, \& J.~Schmidhuber,
\newblock ``Connectionist temporal classification: Labelling unsegmented
  sequence data with recurrent neural networks,''
\newblock {\em Proc. }{\em ICML}, Pittsburgh, 2006.

\bibitem{Zhang2020TransformerTA}
Q.~Zhang, H.~Lu, H.~Sak, A.~Tripathi, E.~McDermott, \etal,
\newblock ``Transformer transducer: A streamable speech recognition model with
  transformer encoders and RNN-T loss,''
\newblock {\em Proc. }{\em ICASSP}, 2020.

\bibitem{jain2019rnntconstrainedbm}
M.~Jain, K.~Schubert, J.~Mahadeokar, C.F.~Yeh, K.~Kalgaonkar, \etal,
\newblock ``RNN-T for latency controlled ASR with improved beam search,''
\newblock {\em arXiv.org}, 1911.01629, 2019.

\bibitem{kurata2020knowledge}
G.~Kurata \& G.~Saon,
\newblock ``Knowledge distillation from offline to streaming RNN transducer for
  end-to-end speech recognition,''
\newblock {\em Proc. }{\em Interspeech}, Shanghai, 2020.

\bibitem{bahdanau2016end}
D.~Bahdanau, J.~Chorowski, D.~Serdyuk, P.~Brakel, \& Y.~Bengio,
\newblock ``End-to-end attention-based large vocabulary speech recognition,''
\newblock {\em Proc. }{\em ICASSP}, Shanghai, 2016.

\bibitem{kannan2018analysis}
A.~Kannan, Y.~Wu, P.~Nguyen, T.N.~Sainath, Z.~Chen, \& R.~Prabhavalkar,
\newblock ``An analysis of incorporating an external language model into a
  sequence-to-sequence model,''
\newblock {\em Proc. }{\em ICASSP}, Alberta, 2018.

\bibitem{kudo2018sentencepiece}
T.~Kudo \& J.~Richardson,
\newblock ``SentencePiece: A simple and language independent subword tokenizer
  and detokenizer for neural text processing,''
\newblock {\em Proc. }{\em EMNLP}, Brussels, 2018.

\bibitem{watanabe2018espnet}
S.~Watanabe, T.~Hori, S.~Karita, T.~Hayashi, J.~Nishitoba, \etal,
\newblock ``ESPnet: End-to-end speech processing toolkit,''
\newblock {\em Proc. }{\em Interspeech}, Hyderabad, 2018.

\bibitem{Panayotov2015LibrispeechAA}
V.~Panayotov, G.~Chen, D.~Povey, \& S.~Khudanpur,
\newblock ``LibriSpeech: An ASR corpus based on public domain audio books,''
\newblock {\em Proc. }{\em ICASSP}, Brisbane, 2015.

\bibitem{park2019specaugment}
D.S.~Park, W.~Chan, Y.~Zhang, C.C.~Chiu, B.~Zoph, \etal,
\newblock ``SpecAugment: A simple data augmentation method for automatic speech
  recognition,''
\newblock {\em Proc. }{\em Interspeech}, Graz, 2019.

\end{thebibliography}
\endgroup
\end{document}